\documentclass[10pt,a4paper]{article}
\usepackage[latin1]{inputenc}
\usepackage[T1]{fontenc}

\title{Values of the Couplings and Internal Geometry}
\author{Amir Muli\'{c} \\ Department of Physics\\
 University of  Oslo \\N-0316 Oslo 3, Norway}

\begin{document}
\maketitle
\begin{abstract}
In this note we display an observation about the geometrical properties of the gauge group manifold of the standard electroweak theory, and the values of the gauge coupling constants. Heuristically obtained structure relates the value 137.036\dots{} to the low energy U(1) coupling. The construction resembles techniques recently used within the framework of the string and D-brane theory. We consider this as an indication that the string theory has direct relevance for the phenomenological QFT and particle physics.
\end{abstract}

This note displays in preliminary way one observation, which possibly indicates that the underlying geometrized theory, for which we shall use the generic name the String Theory (ST), has direct significance for the low energy physics described by the Standard Model (SM) or some its modification. The very simple construction, obtained heuristically, resembles properties of certain existing models and relates the obtained value 137.036\dots{} with the low energy U(1) coupling. We hope that it can shed some light on the nature of the transition between the ST and QFT in general.
\section{Heuristic considerations and motivation}
It is expected, in certain way, that the underlying fundamental theory, if such exists, does not have adjustable free parameters. The values of the couplings in QFT, on the other hand, are not a priori determined at fixed points. E.g. $\alpha_{emIR}$  is  given by experiment, but it can be expected to be, at least in principle, calculable within the some fundamental theory. That is, we expect the fundamental theory to have more degrees of freedom, which are integrated when going to QFT limit and thus provide us with the parameters of the effective theory. We also expect the both theories, fundamental and effective, to have the same infrared behaviour.

	Standard Renormalization Group (RG) arguments would suggest that at low energies one can integrate out all fluctuations of the string except the gauge theory degrees of freedom. This would seem to imply that the ST could not in principle teach us anything about low energy gauge dynamics. Recent work suggests that there are the sectors of ST which are important for the low energy structure of the theory. In brane theory, gauge theory arises as an effective low energy description that is useful in some region in the moduli space of the vacua. Of course, it does not help us much concerning the behaviour at the \emph{literally infrared limit}. Anyway, it is unlikely that without some new ingredient in the theory gauge couplings can be practically calculable. We shall try here to test the applicability of some (at this stage rather loose) ideas to this problem.

	The squared couplings at their places in generating functional really look like the factors normalizing the statistical weights of the field configurations, but within the QFT itself we do not expect some new fundamental degrees of freedom. We can pose the following question. If there is some structure in the underlying geometrized theory responsible for the gauge degrees of freedom, is it possible that it allows a set of models at given energy level, and the measure of this set should be integrated out. Without the specific model, the question seems to be rather vague.

The first key ingredient in our hypothesis is that the manifold with the same geometrical structure as the covering space of the gauge group manifold itself can be used to study the set of possible models. This requires an explanation. Namely, the moduli space of the theory which depends on some parameters is defined as the range of the parameters leading to distinct physics. If the vacua are related by the symmetry, it is not the case, so we cannot use the parts of the gauge group manifold as the moduli space.

To elucidate this issue, let us remind ourselves how the ground state in the electroweak theory was chosen. Using the global gauge freedom , the vacuum condensate was located in the lower, neutral component of the isospinor. It was done using the means of the theory, and not as an actual choice between physically distinct situations. But if in the underlying fundamental theory all possible assignments of the charges (and all possible ways of the symmetry breaking ) consistent with the geometry of the gauge group manifold were allowed, the choice would be a physical one. In matter of fact, we suppose that within the underlying theory the phenomenological gauge charges are not quantized in familiar way.

	We observe that the simple factors of the electroweak gauge group correspond to the non-vanishing cycles on the covering space of the gauge group manifold. The topic of cycles in the internal geometry is a rapidly evolving subject, and an attempt to try something that resembles machinery already in use is the second key ingredient in our hypothesis.

The existence of the solitonic degrees of freedom in the string theory, together with the properties of the internal geometry of the space \textbf{\emph{M}} upon which  the string is compactified leads to many interesting phenomena. In particular \emph{p}-brane can wrap round the \emph{q}-dimensional cycle \emph{C}$\subset$\textbf{\emph{M}} leading to \emph{p-q} dimensional brane in the noncompactified space. The possible \emph{C}'s and their physical consequences has been discussed in numerous contributions, e.g.\cite{Oogu}. Basically, it is possible to consider the vanishing cycles, that are usually considered as a sources of the gauge charges, but we shall be  interested in non-vanishing ones, typically considered in the study of black holes, where also exists, although in a different context, the problem of the "disappeared" degrees of freedom. They can be related to the number of ways the cycle (together with the choice of gauge field on it) can be deformed in the compactification geometry. For an excellent overview of this and related issues see \cite{Vafa}. It should also be mentioned that the models were considered, where strings (WZNW models) and D-branes (models considered e.g. in refs. \cite{Kato,Klim}) live on the group manifolds. Having this in mind, the use of ideas motivated by the research in D-brane theory in the study of  the gauge group manifold does not seem unnatural.
	We are testing our idea in the environment of the electroweak theory. There are three basic reasons to do so. The topology of the gauge group manifold is rather simple, represented by the product of the spheres, charges can be measured with high precision and  inverse coupling is maximal for IR fixed point.
\section{The basic idea and geometrical set-up}
Loosely speaking, we are interested in how many ways the cycle containing the subgroup that corresponds to the particular charge can be embedded in the covering space of the gauge group manifold. To find it out, we need to parametrize the sets of non-vanishing cycles that can support the gauge groups corresponding to the particular charge operators. We consider the cycles that are geometrically similar\footnote{I.e. we are discussing cycles, round spheres etc. Our statements could be expressed e.g. using ``cycles with minimal volume''. At this stage we are trying to avoid mathematical formalization as much as possible.} to the cycles identified as containing the particular subgroup· . After that we shall try to relate the integrals over parameter space to the  phenomenological values of the couplings.
	For the standard electroweak theory, the gauge group manifold is $S^{3}\times S^{1}/Z_{2}$  and we shall use it as a toy model in an attempt to realize this idea.
	Let us first identify such cycles. For the $SU(2)\times U(1)$  and group manifold $S^{3}\times S^{1}/Z_{2}$, with the natural isomorphisms
\begin{equation}
SU(2)\cong S^{3},\;\;U(1)\cong S^1
\end{equation}
any of the points on $S^{3}$ can be identified with the unit matrix and any of the points on $S^{1}$ can be attached to the unit. So any of the $S^{3}$   can be identified  as SU(2) and by this we have identified  as SU(2) also the antipodal $S^{3}$ (in respect to $S^{1}$ ). 
	Having the identification of the two cycles given by (1) , we can see that the cycles that can support the U(1) gauge group belong to the following sets:
\begin{itemize}
\item{}a: set of $S^{3}$  cycles parallel to the  $S^{1}$ that is given by (1) , and in the sense of our statement these cycles can support both SU(2) group and some its U(1) subgroup.
\item{}b: set of  $S^{1}$ cycles normal to  the $S^{3}$  that is given by (1)
\item{}c: cycles  $S^{1}$  described for the given identification (1) by\end{itemize}
\begin{equation}
S^{1} \sim exp(-i\phi )[\;\cos\phi I-i\sin (a_{i}\tau_{i})],\;\;\;a_{i}a_{i}=1 
\end{equation}             
and those  parallel to them.

For $a_{3}=1$    it is obviously the usual  electromagnetism, represented as the subgroup of U(2):
\[\left| \begin{array}{cc}
w&0\\0&1\end{array}\right|,\;\;|w|=1\]                                           
with the generator $\frac{i}{2}(I+\tau^{3})$ .
\begin{itemize}
\item{}	For the case "a" , it is natural to take as the measure of the number of modes the cycle can be located on the gauge group manifold the integral over $S^{1}/Z_{2}$, which equals $\Xi_{a}=\pi$ .
\item{}For the "b" we have in the same sense integral over $S^{3}/Z_{2}$, being equal $\Xi_{b}=\pi^{2}$ .
\item{}For the "c", the situation is slightly more complicated. With  the given point I on the $S^{3}$, every cycle given  by (2) is described by some fixed $\vec{a}$. It is natural to take the integral over the unit sphere $S^{2}$  ($a_{i}a_{i}=1$ from (2)) that equals $4\pi$ as the measure of  such cycles, and this holds for each point on $S^{3}/Z_{2}$ .   So we have $\Xi_{c}=4\pi\;\Xi_{b},\;\Xi_{c}=4\pi^{3}$.\end{itemize}
	In that way, for all possible cycles that can support U(1), including those given by the linear combination of the generators of the simple factors, we have the following measure:
\begin{equation}
\Xi_{a}+\Xi_{b}+\Xi_{c}=137,036\dots
\end{equation}

what is very close to the phenomenological value of the fine structure constant.

If we accept for a moment that we are not dealing with the bizarre coincidence, we are faced with many questions and puzzles. First of all, it is the question about the Weinberg angle. Namely, the value (3) is expected to hold as infrared limit in all possible unbroken low energy $U(1)$ theories. It is not only consequence of the presented construction. It is rather expected property of the theory, see e.g. Ch.18 of the \cite{Polc} for the discussion of the couplings for the continuos set of vacua with unbroken U(1). The very nature of the Weinberg angle is connected with the specific way of the symmetry breaking. The same is truth for the formula from  the electroweak theory
\begin{equation}
\frac{1}{e^{2}}=\frac{1}{g'^{2}}+\frac{1}{g^{2}}
\end{equation}
	                                  
It is not clear how to apply our idea in the cases where the couplings cannot be related to all considered cycles (couplings which are not related to unbroken gauge group). In plain words, (3) is independent of Weinberg angle, contributing only to appropriate normalization of $g$ and $g'$. The only thing that we can do now is to proceed straightforwardly and consider $g^{2}$ and $g'^{2}$ to be  proportional to $1/\Xi_{a}$ and $1/\Xi_{b}$ respectively. As we remember, these are the measures of the sets of cycles  parallel on the $S^{3}\times S^{1}/Z_{2}$ to those identified as weak isospin and hypercharge group. In that way $\sin^{2}\theta_{W} =0.241\dots$ what is roughly satisfactory having in mind that the energy scale has not been precisely identified. Of course that without the elaborated theory seductive numerical result (3) can be in the best case considered only as an indication that we are on the right trace.
\section{The possible meaning of the obtained construction}	
It is premature to link our statements to any specific development of the ST, which was used here only as a motivation. In the same time, we can mention several theoretical constructions that exhibit the properties similar to ones described in this note. First of all, some recent developments in D-brane theory, where  inverse squares of the gauge coupling constants are represented as volumes of the cycles $S^{3}$ and $S^{2}\times S^{1}$  on the  Calabi-Yau manifolds \cite{Oogu}. The volume is determined by the compactification radius
\begin{equation}
\frac{1}{g^{2}}\sim V_{C_{i}}
\end{equation}

The possibility that the infrared couplings correspond to some self dual (in the sense of  T duality) radius deserves further attention, specially because the enhancement of the gauge symmetry is in some cases expected to occur there.

We have already mentioned the branes living on the group manifolds [3].

Strictly speaking, couplings as integrals over the gauge group manifolds and over cycles are known from long ago  in the context of extended field configurations and specially in SUGRA, for recent review see \cite{Town} and references therein, but the concept of duality was not developed enough,  and it was quite unclear how to relate facts about extended configurations to QFT of point-like objects. It is also intriguing that our toy model is completely formally equivalent to the conformally compactified Minkowski space \cite{Witt}, as in the formalism developed by Penrose, what maybe opens the possibility for a link with  holography.

It seems that \emph{in the underlying fundamental theory} the notion of the  charge quantization  must be somewhat modified, at least classically. This would also have consequences for the notions of duality, BPS states and the anomaly structure of the theory which are all conceptually deeply interrelated with the conservation of the charge. 
 
We think that this construction  can be reconciled  with the our present knowledge about the coupling unification. The expected gauge group manifolds at the unification energy scale have different topology and described cycles do not influence directly the values of the couplings, but this issue requires further study.

Presented work has a preliminary and heuristic nature, and it does not make claim about "explanation of $\alpha^{-1}$ " or similar. It rather attempts to help bridge the gap between the phenomenological physics and the ST starting from the phenomenological side. To achieve this, very strong assumptions were made both explicitly and implicitly. In that way we got at once expression (3). This raises the hope that we are on the right trace, but only the future development could justify the proposed identifications.

\section{Acknowledgments}
I would like to thank all people who were, at different stages of the work,  patient enough to discuss presented ideas with me. Of course this does not make any of them responsible for my undue  assumptions.  I specially owe thanks to Kenan Suruliz, Finn Ravndal and Ulf  Lindström.


\begin{thebibliography}{99}
\bibitem{Oogu}H.Ooguri,C.Vafa, Geometry of N=1 dualities in four dimensions, Nucl. Phys.{\bf B} 500 (1997) 62-74 hep-th/9702180
\bibitem{Vafa}Geometric Physics, talk at the ICM98-Berlin hep-th/9810149
\bibitem{Kato}M.Kato, T.Okada, D-Branes on Group Manifolds, Nucl.Phys. {\bf B}499(1997) 583-595 hep-th/9612148
\bibitem{Klim}C.Klimcik, P.Severa, Open strings and D-branes in WZNW models, Nucl. Phys.{\bf B} 488(1997) 653-676 hep-th/9609112
\bibitem{Polc}J.Polchinsky.{\em String Theory}, Cambridge University Press,1998
\bibitem{Town}P.K.Townsend, M-Theory from its superalgebra, Cargese lectures 1997 hep-th/9712004
\bibitem{Witt}E.Witten, Anti De Sitter Space and Holography, Adv.Theor.Math.Physics{\bf 2} (1998)253-291 hep-th/9802150
\end{thebibliography}
\end{document}